# Assessing the Use of Social Media in Massive Open Online Courses

**Suhang Jiang**
University of California, Irvine
Irvine, US
suhangj@uci.edu

**Dimitrios Kotzias**
University of California, Irvine
Irvine, US
dkotzias@uci.edu

**ABSTRACT**
The study explores whether the use of Twitter in Massive Open Online Courses (MOOCs) promotes the interaction among learners. The social network analysis shows that instructors still play a very central role in the social media communication and the communication network between students shrinking over time. The mere use of social media fails to promote learner-learner interaction. More research is needed for understanding learner motivation and how instructional design can help increase their engagement and participation.

**Author Keywords**
Social Media; MOOCs; Social Network Analysis;

**ACM Classification Keywords**
Computer Uses in Education

**INTRODUCTION**
The emergence and development of Massive Open Online Courses (MOOCs) reignited people's interest in online education and pushed it to a new height. Unlike the online education that was provided as a part of an institution's program, MOOCs provide free open online courses to people all around the globe. The leading MOOC platforms such as Coursera, edX, and Udacity offer free online courses taught by professors from elite universities mainly from North America. The courses range from Computer Science, Mathematics to Economics, History. Hundreds of ongoing courses from a wide range of fields are available for people to enroll. The courses usually consist of video lectures, quizzes, weekly assignment and the discussion forum. It is not uncommon for having over 30,000 students enrolled in an online course. For example, one Computer Science course taught by Andrew Ng attracted over 100,000 students to enroll.

Generally, MOOCs differ from the previous online education in three aspects, i.e., learners do not need to register in any institution; learners can access all the courses for free; there is no credit given for completing the MOOC besides the certificate issued by the platforms [1]. Anyone with Internet connection in the world can access the courses provided by Harvard and Stanford. MOOCs expand the free access to quality learning resources on a much bigger scale than traditional online education. It promises to democratize education and provide educational equity for those who otherwise would not be able to receive a quality education.

The nature of MOOCs incurred higher enrollment and also higher attrition rate than traditional online education. MOOCs expanded at an exponential rate since 2011. The major MOOC platforms such as Coursera, edX, and Udacity have attracted over tens of millions of learners to enroll online courses [12]. Nevertheless, it is estimated that in general, the completion rate of MOOCs is less than 7% [16]. Even among learners who intended to complete a course at the beginning, the completion rate is about 22% [20], which is lower than that of traditional online education (about 67%) [27]. The higher attrition rate of MOOCs compared to the traditional online education may be due to the fact that the learners are from more diverse background, with more diverse education experience and motivations than those enrolled in degree-granting institutions, learners have the freedom to take and drop courses without costs and that the certificate issued by the platforms are not widely recognized. Learners reported that the lack of time, insufficient math background and having no intention to complete as the reasons for their early withdraw from the online courses [4].

To tackle the high attrition issue of MOOCs, we draw literature of traditional online education about the factors influencing the success of online education. Social interaction has been suggested as crucial for sustaining learners in traditional online courses [13]. Learners reported that they did not learn well in online courses because they receive less instructor support and encouragement when taking online courses [13]. Previous research shows that the instructor-student interaction influences students' persistence in traditional educational settings. Tinto [24] stated that students are more likely to complete their education the more time the faculty gives to the students. Barnett [3] indicate that instructor's caring, connection, and guidance increases learners' sense of integration, which in turn influences their intent to persist. Therefore, Jaggars [14] suggested that online courses should 'incorporate stronger interpersonal connections and instructor guidance than most currently do'.

Literature from online education and MOOCs also show that social interaction and the sense of community is crucial to the success of an online learning community [18]. Social presence strongly predicts learners' satisfaction of their online learning experience [25]. In addition, social factors have been identified to contribute to attrition in MOOCs [21]. It suggests that social interaction may reduce learners' sense of isolation and increase their engagement. Learners who are more central or more connected in the online learning network tend to stay longer in MOOCs [28].

In addition, there is evidence that social interaction also promotes the performance of learners in online learning. The quality of interpersonal interaction within a course relates positively and significantly to student grades [15]. Learners who play a more central role in the discussion forum have been identified as positively correlated with their cognitive learning outcomes [22]. Research in an online collaborative learning found that central learners tend to have higher final grades and suggested that communication and social networks should be a central element in distributed learning environment [7].

Currently, the discussion forum is the most widely used tool for connecting learners and instructors in online learning settings. It allows learners to connect, exchange ideas and stimulate thinking [6]. However, less than 5% of MOOCs learners ever participated in the discussion forum. The asynchronous communication in the discussion forum does not seem appealing to the majority of learners.

The dilemma of online education needs a paradigm shift as how to organize online teaching and learning effectively. Connectivism was therefore proposed to guide new forms of online education. Connectivism as a hypothesis of learning was first proposed by George Siemens [23] to describe the learning principles and processes that are reflective of the digital age. The metaphor of node and network is central to Connectivism. The key principles of Connectivism include the following: 1) Learning and knowledge rest in diversity of opinions. 2) Learning is a process of connecting specialized nodes or information sources [23].

Thus effective knowledge flow is critical for organizational activities. Downes [10] suggests that Connectivist learning usually consists of four types of activities, i.e., 1) aggregating materials related to the course, 2) remixing and associating materials with each other, 3) creating and practicing, 4) sharing the work. It implies that knowledge is distributed, complicated and individualized [10].

An example of Connectivism is Connectivist MOOCs, i.e. cMOOCs, which usually use social media to allow participants to collaboratively create knowledge. The MOOCs provided in Coursera and edX share similar instructional design patterns with Instructionism as the underlying pedagogy. The courses usually consist of static content, such as video lectures and automatic grading quizzes, and the discussion forums which allow learners to interact and share information. Therefore, Downes came up with the terms "cMOOC" and "xMOOC" to describe the MOOCs underpinned by Connectivism and Instructionism respectively [19]. Despite that both types of MOOCs are freely accessible to the public, it suggests that cMOOCs and xMOOC differ in the following aspects [8]. cMOOCs aim to foster connection and collaboration among learners while xMOOCs emphasize the instructor-learners knowledge transmission. Learners in cMOOCs are expected to contribute to content creation while learners in xMOOCs usually receive information and access content created by instructors.

Under the trend of learning on demand and the wide use of social media, learners are supposed to take an active role of sharing content in the networked world, in addition to merely passively receiving information. Social media is arguably able to facilitate the formation of learning community, promote learner engagement and participation, and overall learning experience for the new-generation learners raised in the 'always-on' world [2]. Learning in the context of social media has become more self-autonomous and informal.

Previous studies investigated the nature of social interaction patterns in a networked learning environment using social network analysis [9]. Nevertheless, the study contexts were mainly about the online programs offered by degree-granting institutions, where learners' population and motivation are more homogeneous than the MOOC learners. In addition, statistical social network analysis was lacking in the previous studies. For example, de Laat and his colleagues mainly used descriptive social network analysis to describe the interaction patterns. This study fills the gap by employing statistical social network analysis, e.g. blockmodelling and conditional uniform graph test, to explore the interaction patterns of MOOC learners on Twitter.

From the Connectivist viewpoint, social network analysis serves an import role in understanding the "learning models in the digital age" [23]. Therefore social network analysis is employed to examine how learners and instructors communicate and interact in the social learning format. The application of social network analysis in educational settings has received considerable momentum in recent years. Lockyer and his colleagues [17] proposed that social network analysis enables researchers and instructors to evaluate the learning design and understand how learners interact with instructors. The instructional design is considered effective only if the communication patterns align with the intended learning design. In the social learning context, learner-centered instead of instructor-centered communication pattern would be considered as achieving the intended design goals.

Therefore the study aims to address the following questions.

1. What is the interaction pattern in the use of Twitter supplementing MOOC?

2. What is the role of instructors in the social media learning?

**DATA**

The data mainly come from the nine-week Data Analytics Learning MOOC, which is offered not only on the edX platform but also in Twitter from Oct 20th to Dec 22nd, 2014. The course was co-taught by four professors from University of Texas, Carnegie Mellon University, Athabasca University, and Columbia University. Each of the instructors covered a topic for two weeks, ranging from prediction modeling, text mining, and social network analysis. The uniqueness of the course lies in that learners are able to customize their learning experience, since they are offered in two formats, i.e., the standard instructor-led format which consists of video lectures and discussion forums, and the social learning path which allows learners to interact in social media such as Twitter, Facebook, and Google group.

Learners can participate in the Twitter social learning path by posting or commenting with the hashtag "dalmooc". The dataset consists of 1,617 tweets which contain the hashtag "dalmooc" during the nine-week course. The 1,617 tweets were produced by 506 users, which compose of 4 instructors and 502 learners. In addition, 1,150 non-missing edges were formed among the 506 users.

We highlight the mentioning interpersonal activity in Twitter in this study. Users interact by mentioning another user with @ username in the tweet. Mentioning is often identified when a user commented or responded to another user. If user A mentioned user B in a tweet, a directed edge from user A to user B is created. Furthermore, we created weekly network graphs in line with the course schedule to track how the communication pattern evolves over time.

**DATA ANALYSIS**

The study mainly employs social network analysis to explore the interaction and communication patterns of learners and instructors in the Twitter social learning path. In the first place, we provide a descriptive analysis on the network graphs. In the second place, node-level centrality and graph-level centralization metrics are analyzed. In the third place, we use inferential tests to test the hypothesis.

**Descriptive Analysis**

To obtain an intuitive understanding of the network, we describe the network size, density and dyad census and visualize the networks. To reflect the different aspects of network size, the study employs the number of nodes, the number of ties. The density of the graphs refers to the proportion of possible ties that are actually present in the graph [26].

Centrality indices are the most common structural indices employed in the analysis of networks. These measures demonstrate the extent to which a node has a central position in the network [11]. Several measures of centrality exist and three of the most common measures were used in this study: degree, betweenness, and closeness. Degree centrality measures the total number of edges to which a node is tied. This represents the number of other users to which one is tied through retweeting or mentioning in Twitter in the study. Those with a high degree have greater levels of participation in a variety of tweets that put them in contact with other users. We also utilize betweenness, which measures the extent to which a node bridges other nodes by lying on a large number of shortest paths between them. Nodes with high betweenness have been described as having some degree of control over the communication of others [11] as well as greater opportunities to exert interpersonal influence over others [26]. Nodes with high betweenness in the social learning path participate in discussions in such a way to bridge other users in the network. Finally, we measure closeness, which measures the extent to which a node has short paths to other nodes in the network. Nodes with high closeness centrality are described as being in the "middle" of the network structure [5]. Analyzing the centrality allows us to understand how central the instructor and learners are in the network graph.

In addition to measuring node-level centrality, the graph-level centralization was analyzed as well. The graph-level centralization indices measure the difference between the most central node and the centrality scores for all other nodes in the network. It provides a graph-level measure of the extent to which centrality is concentrated on a small portion of the network's nodes [26]. We compute these centralization scores for the three aforementioned centrality measures: degree, betweenness, and closeness. These measures demonstrate the extent to which centrality is concentrated on a small number of users.

**Conditional Uniform Graph Test**

The conditional uniform graph (CUG) tests were employed to determine whether features of our observed graph occur at levels exceeding what we would expect by chance. The CUG test conditions on a certain set of network features (e.g., size, the number of edges, or dyad census) and treats all graphs within that set as equally likely. It then draws at random from this set of graphs and measures whether the statistic of interest is greater, less than, or equal to the measure from our original, observed graph. To the extent that few graphs drawn from the set exceed our observed measure, the measure is higher than we expect by chance. In our analyses, we measure whether the observed levels of centralization in the Twitter networks are greater than what we could expect from graphs of the same size with the same number of edges.

## Blockmodel

Blockmodelling produces a simplified representation of the network. It partitions nodes in the network into discrete positions and explores whether ties exist within or between the positions. We used a labeled blockmodel to partition nodes by their node attribute, whether a user is an instructor or a learner. We use a density criterion to create the final block structure [26]. All blocks whose density exceeds the density of the original network are converted to one-blocks while those below the threshold are converted to zero-blocks.

## Assortativity

Assortativity refers to the phenomenon that nodes with the same or similar attributes form ties, which reflects homophily. The assortativity coefficient is the Pearson correlation coefficient of degree between pairs of linked nodes. Positive values of r indicate a correlation between nodes of similar degree, while negative values indicate relationships between nodes of different degree.

## RESULTS

Figure 1 visualizes the social interaction of the 9-week MOOC in Twitter. The four instructors are colored with brown and learners are colored with yellow. The size of the node is proportionate to the degree centrality measurement. It illustrates that an instructor occupies very central positions while the learners are peripheral.

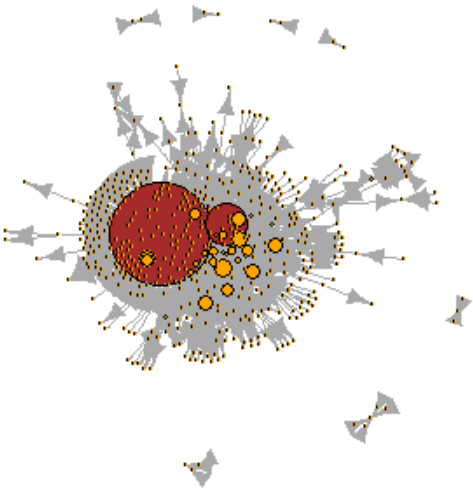

**Figure 1. Whole Graph Visualization**

Table 1 shows how the structure of the social learning community changes over time. The number of nodes and the number of ties is in generally deceasing over time, and the density of the network increases.

| Week | Nodes | Ties | Density |
|---|---|---|---|
| Week1 | 197 | 392 | 0.01 |
| Week2 | 220 | 391 | 0.008 |
| Week3 | 108 | 190 | 0.016 |
| Week4 | 76 | 146 | 0.026 |
| Week5 | 48 | 62 | 0.027 |
| Week6 | 29 | 49 | 0.06 |
| Week7 | 32 | 40 | 0.04 |
| Week8 | 27 | 26 | 0.04 |
| Week9 | 27 | 29 | 0.04 |

**Table 1. Weekly Graph Description**

To determine whether observed graph-level centralization exceeds levels we would expect by chance, we use conditional uniform graph (CUG) tests conditioned on the dyad census. We hold constant the number of nodes and the number of dyads (either mutual or null, given our undirected graph) when running the test. Table 2 shows the result which indicates that that both of our observed networks have much higher levels of centralization than we would expect by chance across the degree and closeness centralization, but not in the betweenness centralization These networks are characterized by concentrations of centrality on a handful of nodes. While certain nodes have high levels of centrality, others lack centrality in the network.

| Week | Degree | Betweenness | Closeness |
|---|---|---|---|
| Week1 | 0.18*** | 0.14 | 0.002*** |
| Week2 | 0.21*** | 0.12 | 0.002** |
| Week3 | 0.28*** | 0.22 | 0.006*** |
| Week4 | 0.29*** | 0.15 | 0.01*** |
| Week5 | 0.17*** | 0.10 | 0.007** |
| Week6 | 0.31*** | 0.16 | 0.02*** |
| Week7 | 0.12*** | 0.09 | 0.007 |
| Week8 | 0.15** | 0.001 | 0.01** |
| Week9 | 0.14** | 0.04 | 0.01* |
| Whole Graph | 0.24*** | 0.15*** | 0.001*** |

**Table 2. Conditional Uniform Graph Test on Centralization**

Notes. * p < .05 ** p<.01 ***p <.001

Blockmodelling is employed to explore the structure of the social interaction. Figure 2 shows the blockmodel that represents the underlying communication pattern of the network. The model suggests that instructors tend to communicate not only with learners but also with other instructors. Nevertheless, the density of learner-learner block fails to exceed the density of the network, which indicates that learners tend to form ties with instructors, but not with other learners.

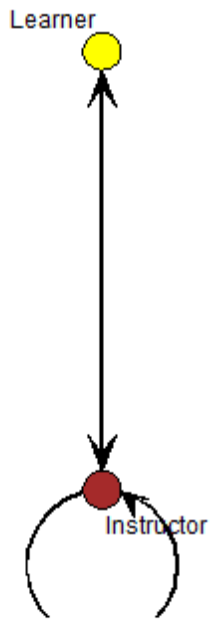

**Figure 2. Learner and Instructor Interaction Pattern**

In addition, we examine the assortativity in terms of a node's attribute (learner or instructor) and the value is -0.12, which indicates that learners tend to connect with instructors and vice versa, but not within learners. The assortativity measures the level of homophily of networks, based on the node attribute whether he or she is instructor or learner. The homophily negative assortativity coefficient indicates that dissimilar nodes tend to connect with each other. It suggests that the social learning network does not possess identity homophily.

## CONCLUSION

The result shows that the use of Twitter fails to promote learner-learner interaction, and the instructor-led communication pattern still dominates. The cMOOCs which use social media to engage learners and customize their learning experience did not promote the expected learner-learner interaction. Similar to xMOOCs, cMOOCs also experience significant dropout as time goes by. Simply using social media cannot fundamentally increase learner interaction. We propose that more research is needed to understand learner motivation and how the instructional design can help engage learners.

## ACKNOWLEDGMENTS

We are very grateful for the support of the Digital Learning Lab, University of California, Irvine.